# A Tool to Facilitate Web-Browsing

Christopher A. Kelly[^,1,2,3,4,5] Jonatan Fontanez[3,6] & Tali Sharot[^,1,2,3]

[1]Department of Experimental Psychology, University College London, London WC1H 0AP, UK
[2]Max Planck University College London Centre for Computational Psychiatry and Ageing Research, London WC1B 5EH
[3]Department of Brain and Cognitive Sciences, Massachusetts Institute of Technology
[4]Institute for Human-Centered AI, Stanford University, Stanford, CA, USA.
[5]Department of Psychology, Stanford University, Stanford, CA, USA.
[6]Department of Electrical Engineering and Computer Science, Massachusetts Institute of Technology

[^]Corresponding authors: cakelly@stanford.edu, t.sharot@ucl.ac.uk

## Abstract

Search engine results often misalign with users' goals due to opaque algorithms, leading to unhelpful or detrimental information consumption. To address this, we developed a Google Chrome plugin that provides "content labels" for webpages in Google search results, assessing Actionability (guiding actions), Knowledge (enhancing understanding), and Emotion. Using natural language processing and machine learning, the plugin predicts these properties from webpage text based on models trained on participants' ratings, effectively reflecting user perceptions. The implications include enhanced user control over information consumption and promotion of healthier engagement with online content, potentially improving decision-making and well-being.

# Overview

Approximately eight billion search engine queries are submitted daily by individuals who seek to gain knowledge and make informed decisions (Kemp, 2022). However, search results are shaped by opaque algorithms that do not necessarily align with users' goals (Rainie, Lee & Anderson, 2017). Consequently, individuals dedicate countless hours to absorbing information that may not yield practical benefits, and in some cases, may have a detrimental effect on their well-being (Kelly & Sharot, 2023). For example, by consuming negatively valanced information that is not informative or helpful.

To address this problem we developed a tool designed to empower users to navigate the web in a way that may improve their decision making, mental health, and understanding. Much like how people use nutritional labels to learn about the nutritional value of food before it enters their body (e.g., calories, fat content etc.), the tool provides 'content labels' for available webpages in a search engine results page that a user can inspect before consuming information.

In particular, the software (in the form of a Google Chrome plugin) informs users of three properties that can guide information-consumption decisions: (i) actionability (the ability of text on a webpage to guide action, on average); (ii) ability of text on a web page to enhance understanding, on average; (iii) sentiment (e.g., how positive or negative the text on a webpage is). These three properties were selected based on empirical research that indicates that people's key motives for seeking information is to (i) guide their actions and decisions (Kelly & Sharot, 2021; Stigler, 1961; for review Sharot & Sunstein, 2020), (ii) improve comprehension (Sharot & Sunstein, 2020) and (iii) improve affect (Kelly & Sharot, 2021; Sharot & Sunstein, 2020; Charpentier et al., 2018; Loewenstein, 1994; Caplin & Leahy, 2001; Kőszegi, 2010; Golman et al., 2017).

The plugin provides scores visible in a Google search results (**see Figure 1**) about the above three factors [which we respectively called 'Actionability', 'Knowledge', and 'Emotion'] of text found on webpages. Users can use these scores to improve their web-browsing experience, such that the information they consume better aligns with their goals. For instance, individuals seeking practical advice such as *"I just lost my job"* may prioritise information with a high 'Actionability' value, while those looking to deepen their understanding of a topic, for example *"who is the most famous pharaoh"* might prioritise webpages with a high 'Knowledge' score.

Indeed, different individuals may prioritise some of these scores over others (Kelly & Sharot, 2021). For example, some people may be driven more to seek information that can help them make better decisions, while other may be primarily driven to seek information that helps them understand the world better. The importance of these motives can vary as a function of a person's state (e.g., stress vs. relaxed etc.; see Kelly & Sharot, 2024) and domain (for example, in the domain of health 'Actionability' of information may be especially important; Kelly & Sharot, 2021).

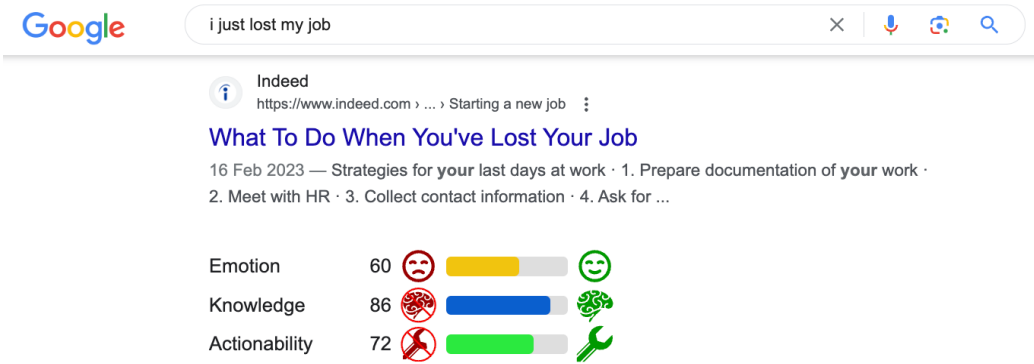

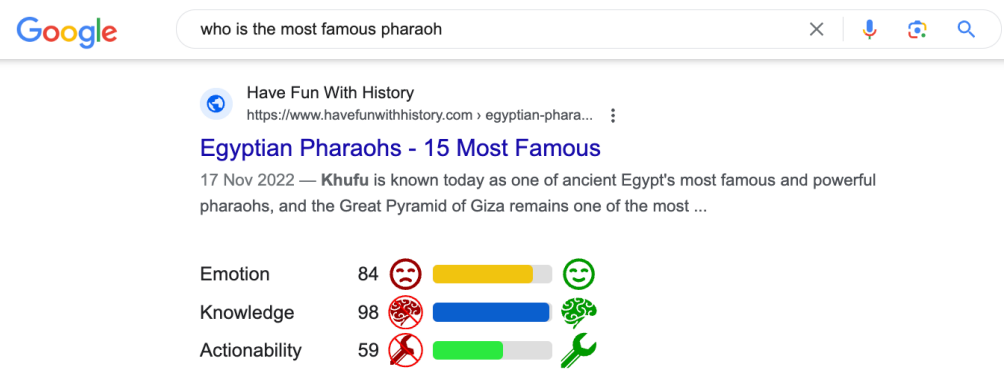

**Figure 1. Presentation of Scores.** The figure presents the scores for webpages obtained from the Google search engine for two different queries: **(a)** *"I just lost my job"* and **(b)** *"who is the most famous pharaoh".* The *Emotion* (yellow), *Knowledge* (blue), and *Actionability* (green) scores are computed for each webpage listed in the Google search results using the process described in the **Tool Development** section below. The user is then presented with these scores alongside each webpage. This feature enables the user to make informed decisions about which webpage to visit, which can improve their web-browsing experience.

The nature of how webpages are interpreted and rated is of course subjective. For instance, a webpage that is perceived as positive by one person may be perceived negatively by another. Yet, as the results detailed below demonstrate, there is nonetheless high agreement across users ***on average*** regarding the emotion, actionability and potential knowledge enhancement of webpages. This suggests that despite subjectivity and individual differences, it is possible to effectively capture a shared perception that is relevant to many users and can be leveraged. Just as mean ratings of products (books, movies, items) are often helpful despite their subjective nature, 'on average' scores of websites can be valuable in guiding users' online information consumption, allowing them to engage with information that aligns with their goals and preferences.

## Tool Development

To measure and present the scores of interest along the Google search results we applied the following method (**see Figure 2**):

**Webpage Retrieval & Parsing:** For each Google search submitted by users, we extracted the base HyperText Markup Language (HTML) source code from each of the web pages and parsed the code using the Python package 'beautifulsoup4' (Richardson, 2007). We then extracted the paragraph and header text from each of those web pages.

**Actionability, Knowledge & Emotion Scoring:**

*Participant Ratings and Label Collection*
We recruited 1499 participants via Prolific's online platform (www.prolific.com). Each participant was instructed to browse and rate five webpages on two dimensions: Actionability, Knowledge, resulting in 7495 total ratings per dimension. We successfully extracted text from 6443 webpages, which was then used for model training and validation (details below). Additionally, 500 participants rated the Positive and Negative Emotion of the webpages, resulting in 2500 ratings for each. Text was successfully extracted from 2174 of these rated webpages. Participants were compensated at a rate of £9.00 per hour. Ethical approval was obtained from the Research Ethics Committee at UCL, and all participants provided informed consent.

Actionability was defined as the extent to which the information on the webpage could guide actions and/or decisions ("*Could the information on the webpage help guide actions and/or decisions?*"). Knowledge was defined as the degree to which the information on the webpage increased the participant's understanding of the topic ("*Does the information on the webpage increase your understanding of the topic?*"). Positive and Negative Emotion were defined as the degree to which the information on the webpage was positive or negative in emotion ("*How Positive is the information on the webpage?*"; "*How negative is the information on the webpage?*"). All dimensions were rated on a 6-point scale, with 1 representing the lowest level and 6 the highest. An overall Emotion score for each webpage was computed by subtracting the Negative score from the Positive score, which we used in our analysis.

*Model Training and Evaluation*
Whether information will help guide a person's action, increase their understanding of a topic or is perceived as positive/negative will obviously alter from person to person. However, it is possible that **on average** some webpages contain information that is more likely to guide a person's action and/or increase their understanding on the topic or is perceived as positive. To test whether there is a good relationship between the models predictions and participants actual ratings of webpage content, we implemented the following natural language processing and machine learning pipeline.

*Data Preprocessing*
Rows containing popup text (e.g., 'cookies', ''blocked') or error messages (e.g., '403', '404', 'error') were excluded. The text was cleaned by removing special characters and extra whitespace to ensure consistency.

*Embedding Generation*
We used the RoBERTa (Robustly Optimized BERT Pretraining Approach) model to generate contextual embeddings for the webpage text. RoBERTa is a transformer-based model known for its effectiveness in understanding language context and semantics. The embeddings obtained from RoBERTa were standardized using a StandardScaler to normalize the feature set, which is essential for models sensitive to the scale of data.

*Data Splitting*

To prevent data leakage and ensure independence between training and testing sets, we split the data based on unique participant IDs using a GroupShuffleSplit. This ensured that all ratings from a specific participant were confined to either the training set or the test set. The data was split into 80% for training and 20% for testing, maintaining the distribution of the target variables.

*Model Development*

Using the XGBoost regression algorithm, we performed an exhaustive grid search with cross-validation (GridSearchCV) to identify the optimal hyperparameters for the XGBoost model. The parameters tuned included:
- Number of Estimators: 50, 100, and 200.
- Learning Rate: 0.01, 0.1, and 0.2.
- Maximum Depth: 3, 5, and 7.

The performance metric used to evaluate the model was the Pearson correlation coefficient between the model predictions and actual ratings on the test set. The grid search identified the best model with the following parameters: number of estimators = 200, learning rate = 0.1, and maximum depth = 5. For the Actionability dimension, the model achieved an r value of 0.41, $p < 0.001$, on the test set. For the Knowledge dimension, the model achieved an r value of 0.58, $p < 0.001$, on the test set. Finally for the Emotion dimension, the model achieved a r value of 0.61, $p < 0.001$ on the test set. These Pearson correlation coefficients suggest that the model is reasonably effective at capturing the relationship between the textual content of webpages and the Actionability, Knowledge and Emotion ratings.

**Storage & Presentation of Values:** The computed Emotion, Actionability, and Knowledge scores were stored in a system database. These scores can be subsequently distributed to users, system tools (e.g., browser plugins), or third parties (e.g., search engines). To maintain up-to-date scores, the process can be repeated periodically whenever the webpage content changes, its formatting is altered, or on a recurring interval basis (e.g., daily).

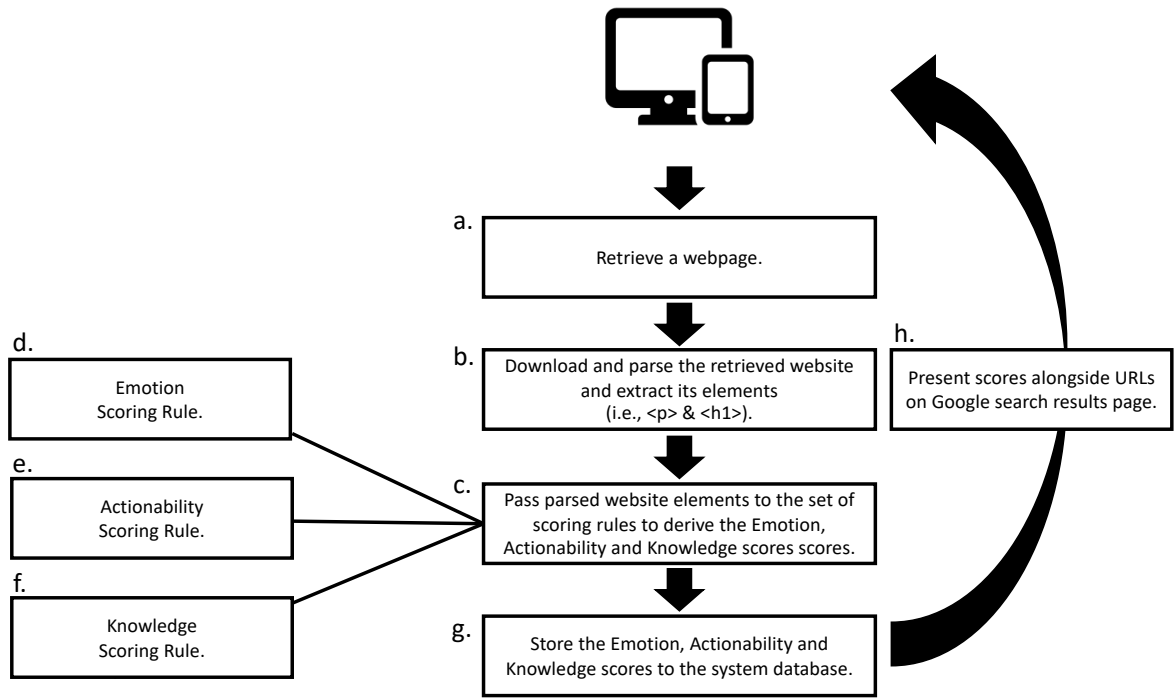

**Figure 2. Process of Tool.** The figure shows a visual representation of the tool's process. **(a)** the URLs of the webpages that users are exposed to on a Google search results page are retrieved. **(b)** Then, the HTML header and paragraph text of the webpages was downloaded and prepared for analysis. **(c-e)** The scoring rules for Emotion, Actionability, and Knowledge are defined and applied to the text. **(g)** The computed scores are stored in a database for subsequent distribution to users via a plugin. **(h)** The plugin presents the scores to users in real time enabling users to make informed decisions and adjust their information-consumption tendencies.

**Non-Model Considerations:** When scraping text from webpages, some considerations must be taken into account due to the varied nature of all webpages on the internet and internet security concerns.

To account for the variation of webpages, we added both a minimum and maximum word count to input into the model. This word count does not include stop words removed during pre-processing. If the list of words scraped from the webpage's HTML is less than 10 words long, the algorithm determines it is too few words and returns invalid values instead of trying to calculate the scores. If the list of words is more than 200 words long, the algorithm stops processing text and passes the words into the model to calculate the scores.

When scraping from certain webpages which have anti-scraping measures, the request can return an HTML that does not reflect the information on the webpage itself. For example, with pages protected by Cloudflare anti-scraping measures, the HTML has text like "*This website is using a security service to protect itself from online attacks. The action you just performed triggered the security solution*". If the model were to be run on texts like this, the scores calculated would be similar between webpages with extremely different content. Text like this is recognized and invalid values are returned.

One simple spoofing technique we can apply is adding an HTML request header. This header provides some metadata to the receiver of the request. By adding information that a real user would have when making a real request to the webpage, such as the browser name and version they are using, some anti-scraping measures can be bypassed.

# Discussion

In the digital era, search engines have become an integral part of information gathering, with billions of queries submitted daily. Despite their prevalence, traditional search algorithms often fail to align with users web-browsing goals. To bridge this gap, we developed a unique tool, akin to nutritional labels for web content, to empower users to make informed decisions about the information they consume online.

Our tool, realised as a Google Chrome plugin, applies natural language processing (NLP) techniques to assign 'Actionability', 'Knowledge', and 'Emotion' scores to webpages. We based these criteria on empirical research suggesting these properties are pivotal to users' information-seeking motives (Sharot & Sunstein, 2020; Kelly & Sharot, 2021). These scores are prominently displayed in Google search results (**see Figure 1**), guiding users in their web-browsing journey to align the information consumed more closely with their individual goals and preferences.

In terms of performance, the Pearson correlation coefficients were moderate to good for all three metrics, demonstrating the tool's effectiveness in classifying webpages according to their 'Actionability', 'Knowledge' and 'Emotion' properties. However, more data is needed to improve the models performance. Together, these results suggest that despite the subjectivity and individual differences in interpretation, our tool was able to capture a shared perception. The scores can thus offer valuable guidance for users' online information consumption, allowing them to engage with information that aligns with their goals and preferences.

In the pursuit of continuously enhancing user experience and the tool's functionality, we have several potential directions to explore:

> 1**. Sort by Function:** We propose to allow users to reorder search results according to their preferred metrics. For example, presenting links ranked from the most actionable to the least actionable. Such a feature could add an extra layer of customisability and empower users to tailor their information exposure according to their needs or preferences.
>
> 2. **Filter by Function:** Building on the 'Sort by Function', we suggest incorporating a filtering mechanism that allows users to eliminate search results based on one or multiple scores. Users might, for example, wish to exclude links with Knowledge and Actionability scores equal to or below 20. This approach could also be adapted as a parental tool, helping to guide children's online exposure.
>
> 3. **Track Web-Browsing Patterns Over Time:** Similar to apps that track physical activity or calorie intake, we envisage a feature that allows users to monitor their web-browsing patterns in relation to the three scores over time. This could provide valuable insights into their information-consumption tendencies and offer them the opportunity to adjust their browsing habits accordingly.

These proposed enhancements aim to augment user control over their web browsing, which may promote healthier, more constructive engagement with online content.

In addition, our current plugin focuses solely on analysing text and does not assess images and videos. While the results obtained suggest text analysis effectively reflects users' webpage ratings, we aim to broaden the tool's capabilities and include a diverse range of media types, such as images, videos, and other multimedia formats. By embracing these varied forms of media analysis, our goal is to create a more comprehensive and powerful tool that can assess all webpages and offer users a richer web-browsing experience.

The next step is to make the tool available to a diverse group of subjects to test (i) whether people use the tool (i.e. does exposure to the scores lead to changes their web-browsing patterns) and (ii) whether using the tool improves people's mood, subjective sense of knowledge enhancement and sense of empowerment.

To conclude, our tool provides a novel solution to the shortcomings of traditional search algorithms by equipping users with an intuitive scoring system to assess web content. By integrating user-driven properties into search results, it enhances the browsing experience and facilitates more goal-oriented and effective online information consumption.